# ON THE PROBLEM OF AMBIGUITY OF ELECTROMAGNETIC POTENTIAL

A.V. Gritsunov
Kharkiv National Economical University
Address: 9a, Lenin Ave., Kharkiv, 61166, Ukraine
Tel. +380 95 0003502, E-mail gritsunov@gmail.com

**Introduction**

Theory of electromagnetism is one of the most accomplished parts of the physics and engineering. This is demonstrated by successful solutions of many important technical problems with this science. However, new progress in technology, especially nanotechnology development, reanimate some "chronic" fundamental problems of the electromagnetism like to the Aharonov-Bohm effect [1] interpretation from the position of the locality principle; the problem of the electron rest mass; vagueness of the magnetic monopole existence, *etc*.

A problem of the electromagnetic potential reality is suspended for a many years. The principal argument "contra" is that any Lorentz-invariant gauge of the potential four-vector $\vec{A}(t,x,y,z) = \{A_t, A_x, A_y, A_z\}$ [2] does not affect the Lagrange equations, total energy and momentum of a closed electromagnetic system in the field formalism. The reason is that all those expressions depend on the electromagnetic field tensor [2], which do not vary by the Lorenz gauge. A value that is not defined unambiguously cannot be treated as real.

On the other hand, the field formalism is unable to explain satisfactorily some recent experiments in the electromagnetism, e.g., the Aharonov-Bohm effect. The locality principle cannot be fulfilled if the charged quantum particle wave function is considered as depending on electromagnetic field intensity in some distant area. Because a self-sufficient potential formalism [3], treating all electromagnetic phenomena as natural or forced oscillations of some distributed electromagnetic oscillating system (Minkowski space-time), was recently offered, let's consider what new this can bring in the problem of potential gauge ambiguity.

**Main Part**

The Lagrange function for a system of charged particles coupled only with the electromagnetic interactions is postulated in the potential formalism [3] using so-called "gradient" hypothesis as

$$\Lambda(t) = \Lambda^P(t) + \Lambda^I(t) + \Lambda^S(t) = -c^2 \sum_{n=1}^{N} m_{0n} \sqrt{1 - (dx_n/dt)^2 - (dy_n/dt)^2 - (dz_n/dt)^2} - \int_V \vec{A} \cdot \vec{j}\, dxdydz - \frac{1}{2\mu_0} \int_V \left[ (\vec{\nabla} A_t)^2 - (\vec{\nabla} A_x)^2 - (\vec{\nabla} A_y)^2 - (\vec{\nabla} A_z)^2 \right] dxdydz , \quad (1)$$

where $V$ is the 3D volume of the system; $N$ is the number of particles with rest masses $m_{0n}$ and charges $q_n$ ($n = 1, 2, \ldots, N$) in this system; $x_n(t), y_n(t), z_n(t)$ are coordinates of the particles; $\vec{j}(t,x,y,z)$ is the current density four-vector; the dot means scalar multiplication of four-vectors: $\vec{a} \cdot \vec{b} = a_t b_t - a_x b_x - a_y b_y - a_z b_z$; $\vec{\nabla} = \{\partial/\partial t, -\partial/\partial x, -\partial/\partial y, -\partial/\partial z\}$ is the four-gradient operator; $c$ is the light velocity; $\mu_0$ is the permeability of free space. $t$ is a time coordinate in meters defined as the product of time and $c$. $\Lambda^P(t)$ is a "mechanical" component for the non-interacting particles; $\Lambda^I(t)$ is an "interaction" component coupling the particles with the oscillating system; $\Lambda^S(t)$ is a component for the distributed oscillating system (Minkowski space-time) itself.

The energy-momentum density tensor [2] $[w] = [w^P] + [w^I] + [w^S]$ is derived from (1) as consisting of the same three "partial" tensors. In particular, the "interaction" energy-momentum density tensor is

$$[w^I] = \begin{bmatrix} A_t j_t & A_t j_x & A_t j_y & A_t j_z \\ A_x j_t & A_x j_x & A_x j_y & A_x j_z \\ A_y j_t & A_y j_x & A_y j_y & A_y j_z \\ A_z j_t & A_z j_x & A_z j_y & A_z j_z \end{bmatrix}. \quad (2)$$

The distributed electromagnetic oscillating system energy-momentum density tensor components are:

$$w_{\tau\tau'}^S = -\frac{g_{\tau\tau}g_{\tau'\tau'}}{\mu_0}\left(\frac{\partial A_t}{\partial \tau}\frac{\partial A_t}{\partial \tau'} - \frac{\partial A_x}{\partial \tau}\frac{\partial A_x}{\partial \tau'} - \frac{\partial A_y}{\partial \tau}\frac{\partial A_y}{\partial \tau'} - \frac{\partial A_z}{\partial \tau}\frac{\partial A_z}{\partial \tau'}\right) + \frac{g_{\tau\tau'}}{2\mu_0}\left[(\vec{\nabla}A_t)^2 - (\vec{\nabla}A_x)^2 - (\vec{\nabla}A_y)^2 - (\vec{\nabla}A_z)^2\right], \qquad (3)$$

where $\tau$ and $\tau'$ are generic symbols meaning any of the coordinates $t, x, y, z$; $[g]$ is the Minkowski space-time unit metric tensor [2].

The purpose of this paper is recognition of the limits for the electromagnetic potential gauge ambiguity in the potential formalism. The gauge ambiguity consists in permissibility of adding to $\vec{A}$ some another four-vector $\vec{\nabla}f$, where $f(t,x,y,z)$ is a relativistic scalar function, and can be expressed in the mathematical form as $\vec{\nabla} \times (\vec{A} + \vec{\nabla}f) \equiv \vec{\nabla} \times \vec{A}$, where an antisymmetric tensor

$$\vec{\nabla} \times \vec{a} = \begin{bmatrix} 0 & \frac{\partial a_t}{\partial x} + \frac{\partial a_x}{\partial t} & \frac{\partial a_t}{\partial y} + \frac{\partial a_y}{\partial t} & \frac{\partial a_t}{\partial z} + \frac{\partial a_z}{\partial t} \\ -\frac{\partial a_x}{\partial t} - \frac{\partial a_t}{\partial x} & 0 & \frac{\partial a_x}{\partial y} - \frac{\partial a_y}{\partial x} & \frac{\partial a_x}{\partial z} - \frac{\partial a_z}{\partial x} \\ -\frac{\partial a_y}{\partial t} - \frac{\partial a_t}{\partial y} & \frac{\partial a_y}{\partial x} - \frac{\partial a_x}{\partial y} & 0 & \frac{\partial a_y}{\partial z} - \frac{\partial a_z}{\partial y} \\ -\frac{\partial a_z}{\partial t} - \frac{\partial a_t}{\partial z} & \frac{\partial a_z}{\partial x} - \frac{\partial a_x}{\partial z} & \frac{\partial a_z}{\partial y} - \frac{\partial a_y}{\partial z} & 0 \end{bmatrix}$$

is interpreted as a four-curl of the four-vector $\vec{a}$. An addition condition $\vec{\nabla} \cdot \vec{A} = 0$ is imposed by obligatory for any relativistic-invariant system Lorenz gauge. As $\vec{\nabla} \cdot \vec{\nabla} = \nabla^2$, this results in a homogeneous wave equation

$$\nabla^2 f = 0. \qquad (4)$$

$\vec{\nabla} \cdot \vec{a} = \partial a_t / \partial t + \partial a_x / \partial x + \partial a_y / \partial y + \partial a_z / \partial z$ is a four-divergence of $\vec{a}$; $\nabla^2 = \partial^2 / \partial t^2 - \partial^2 / \partial x^2 - \partial^2 / \partial y^2 - \partial^2 / \partial z^2$ is the D'Alembertian.

Arbitrary choice of the function $f$ is valid until this choice does not affect any physically objective effect or value. In particular, the above gauge ambiguity hold in the field formalism only so far as the Lagrange equation for a classic charged particle and the expressions for the field energy-momentum density tensor components are based on the four-curl $\vec{\nabla} \times \vec{A}$ (called as "electromagnetic field tensor" in the field formalism). It can be shown that the wave function of a quantum charged particle also is "insensitive" to the Lorenz gauge [4].

However, the expressions (2) and (3) for the energy-momentum density tensor in the potential formalism are not founded on $\vec{\nabla} \times \vec{A}$ tensor. Therefore, the bounds of the electromagnetic potential gauge ambiguity in the potential formalism must be checked on the condition that this ambiguity does not affect the closed (insulated) electromagnetic system total energy-momentum four-vector $\vec{P} = \{P_t, P_x, P_y, P_z\}$. Moreover, the interpretation of the electromagnetic potential as a real physical variable imposes one more constraint on the gauge: all components of $\vec{A}$ must be finite for the whole Minkowski space-time. This excludes linear (with a non-zero coefficient) and exponential (with a real argument) dependences of $\vec{\nabla}f$ components on $\tau$.

Generally, the concept of a "closed" or "insulated" system, having negligibly small interactions with other systems in comparison with the internal interactions, differs for the field and the potential formalisms. In the first, a system containing no charged particles interacting distinctly with electromagnetic fields of other systems can be called as "closed". E.g., a single charged particle within a closed perfectly conductive cavity having permanent charge or a charged particle nearby an infinitely long solenoid with invariable current are insulated systems from this point of view. The rest mass of such systems $M_0$ can be defined from the equation:

$$c^2 M_0^2 = (\vec{P})^2. \qquad (5)$$

On the contrary, the potential formalism defines insulated system as containing no charged particles which are affected noticeably by "disturbances" of electromagnetic potential produced by other systems (i.e., deviations of $\vec{A}$ from the "undisturbed" state when both natural and forced electromagnetic oscillations of the space-time are absent). The abovementioned particles within the closed cavity and nearby the infinitely long solenoid are not insulated systems in the potential formalism, because $A_t$ and $A_\xi$ components respectively for the areas including those particles differ from the ones in the cases of absence of the cavity charge and the solenoid current respectively. This is inadmissible to calculate the rest masses of such particles under (5). Only total masses of the systems "the cavity with the charged particle inside it" and "the solenoid with the charged particle nearby it" can be defined. Note that "insulation" of an electromagnetic system does not suppose the electric neutrality of one. It is enough if the electromagnetic fields (or the potential disturbances) produced by this system do not reach for other systems distinctly.

Analyzing possible solutions of (4) (these can be easily obtained with the variable separation method) on the condition that all components of $\vec{\nabla}f$ are finite for the whole space-time, it can be found that only two types of $\vec{A}$ gauge are formally valid:

(i). Adding arbitrary finite constant to any component of $\vec{A}$.

(ii). Adding components of "*Zero Magnetic*" (*ZM*) natural oscillation of the potential [4] to $\vec{A}$:

$$\left(\vec{\nabla}f\right)_t = A_0 \exp(ik_{et}t)\exp(-ik_{ex}x)\exp(-ik_{ey}y)\exp(-ik_{ez}z);$$

$$\left(\vec{\nabla}f\right)_\xi = A_0 \frac{k_{e\xi}}{k_{et}}\exp(ik_{et}t)\exp(-ik_{ex}x)\exp(-ik_{ey}y)\exp(-ik_{ez}z)$$

[except for $k_{et} = k_{ex} = k_{ey} = k_{ez} = 0$ that is a part of (i)], where $k_{et}$ and $k_{e\xi}$ are components of the four-wave-number $\vec{k}_e = \{k_{et}, k_{ex}, k_{ey}, k_{ez}\}$, $(\vec{k}_e)^2 = k_{et}^2 - k_{ex}^2 - k_{ey}^2 - k_{ez}^2 \equiv 0$ for the natural oscillation; $A_0$ is arbitrary constant; $\xi$ is a generic symbol meaning any of the coordinates $x, y, z$.

It is obviously from (2) that the total energy and momentum of a closed system are depending on the gauge (i) if the total charge $\sum_{n=1}^{N} q_n$ of this system is not equal to zero (because of the "interaction" components changing). Thus, cannot be ambiguity for this type of gauge. This is evidently for elementary physical reasons (e.g., the equality of energies and momentums of an alone charged particle in the field and the potential formalisms, or the equality of the rest masses of charged particles and their antiparticles) that for the "undisturbed" electromagnetic state of the Minkowski space-time all components of $\vec{A}$ must be equal to zero.

The gauge (ii) also results in the same reasoning. Moreover, any electromagnetic system subjected to the *ZM* oscillation cannot be considered as "insulated".

The only rational explanation can be offered for the described above: the electromagnetic potential four-vector is a relative coordinate, not "absolute", for the oscillating system. Namely, $\vec{A}$ four-vector may be physically construed as a 4D deviation of the distributed electromagnetic oscillating system (Minkowski space-time) from its "undisturbed" state. Therefore, all components of $\vec{A}$ are identically equal to zero in the "undisturbed" state, e.g., at a large distance from all free charges and currents.

## Conclusions

The electromagnetic potential must be considered as some relative measure describing deviation of the distributed electromagnetic oscillating system (Minkowski space-time) from the "undisturbed" state (when both natural and forced oscillations are absent). Therefore, there is no ambiguity in the gauge of one: all components of the potential four-vector are asymptotically verging towards zero while the distance from all free charges and currents enlarges. Such interpretation turns the electromagnetic potential into a physically real value.

## References

1. Y. Aharonov, and D. Bohm, "Significance of Electromagnetic Potentials in Quantum Theory", Physical Review, Vol. 115, No. 3, pp. 485-491, 1959.
2. L.D. Landau, and E.M. Lifshitz, "Course of Theoretical Physics," The Classical Theory of Fields, Vol. 2, Butterworth-Heinemann, Oxford, 1975.
3. A.V. Gritsunov, "Self-sufficient Potential Formalism in Describing Electromagnetic Interactions", Radioelectronics and Comm. Systems, Vol. 52, No. 12, pp. 649-659, 2009.
4. A.V. Gritsunov, "On the Reality of "Zero Magnetic" Oscillations of Potential", 2012 Int. Vacuum Electronics Conf. (IVEC 2012), Monterey, CA, pp. 409-410, 2012.